\documentclass[manuscript]{aastex631}

\shorttitle{INTENSITY OF HOMOLOGOUS SOLAR JETS}
\shortauthors{Yang et al.}

\graphicspath{{./}{figures/}}

\begin{document}

\title{Intensity of Homologous Solar Jets: Late-stage Magnetic Cancellation Triggers Major Jet with Remarkable Drift Motion}

\author[0000-0003-3462-4340]{Jiayan Yang}
\affiliation{Yunnan Observatory, Chinese Academy of Sciences, P.O. Box 110, Kunming 650011, People's Republic of China;
yjy@ynao.ac.cn}
\affiliation{Yunnan Key Laboratory of Solar Physics and Space Science, Kunming 650011, People's Republic of China}

\author{Junchao Hong}
\affiliation{Yunnan Observatory, Chinese Academy of Sciences, P.O. Box 110, Kunming 650011, People's Republic of China}
\affiliation{Yunnan Key Laboratory of Solar Physics and Space Science, Kunming 650011, People's Republic of China}

\author{Bo Yang}
\affiliation{Yunnan Observatory, Chinese Academy of Sciences, P.O. Box 110, Kunming 650011, People's Republic of China}
\affiliation{Yunnan Key Laboratory of Solar Physics and Space Science, Kunming 650011, People's Republic of China}

\author{Yi Bi}
\affiliation{Yunnan Observatory, Chinese Academy of Sciences, P.O. Box 110, Kunming 650011, People's Republic of China}
\affiliation{Yunnan Key Laboratory of Solar Physics and Space Science, Kunming 650011, People's Republic of China}

\begin{abstract}

Using high-resolution observations from the New Vacuum Solar Telescope and {\it Solar Dynamics Observatory}, we analyzed a group
 of homologous jets triggered by the emergence of a magnetic bipole in the positive following-polarity region of active region 
12781 on 2020 November 11. The negative polarity of the bipole moved northwestward after its emergence and then cancelled with 
the surrounding pre-existing positive field. The evolution of the negative flux exhibited three distinct stages: emergence, 
stabilization, and cancellation, each producing jets with markedly different properties. The emergence stage produced numerous but 
mild jets, the stable stage yielded the fewest and weakest jets, while the cancellation stage triggered the strongest jet activities. 
Additionally, the most violent jet, J4, occurred shortly after rapid cancellation. J4 exhibited the brightest jet bright point (JBP) 
and a clear lateral drift motion of its axis. The JBP moved in the same direction as the negative polarity, indicating that the 
reconnection site slid in response to the bipole's motion. J4's drift velocity was approximately one order of magnitude higher than 
that of the bipole and that of the JBP. Our analysis suggests that the most intense magnetic free energy release in these 
homologous jets occurred late in flux cancellation, and that J4's drift motion results from the combined effect of whip-like 
motion induced by the sling-shot process and sliding of the magnetic reconnection site, with the former dominant.

\end{abstract}

\keywords{Solar activity (1475) --- Solar magnetic fields (1503) --- Solar corona (1483)}

\section{Introduction} \label{sec:intro}

Solar jets are ubiquitous transient phenomena in the solar atmosphere \citep{raouafi16, shen21}. They are collimated, beam-like 
plasma ejections that occur in various regions of the solar surface, including active regions (ARs) \citep{hong16, yang19}, quiescent 
regions \citep{yang24}, coronal holes \citep{chen12}, and others. Previous studies suggest that approximately 68\% of jets originate 
in ARs \citep{joshi21}. Jets have been observed across a wide range of temperatures and wavelengths, from $H\alpha$ \citep{roy73, 
jiang07, chen08} to extreme-ultraviolet (EUV) \citep{hong17, yang23a} and X-ray wavelengths \citep{shibata92, nistico09, tang21}. 
Accumulating evidence indicates that these small-scale activities could contribute significantly to the energy and mass budgets of 
the solar corona and the solar wind. 

It is generally accepted that the generation of solar jets is closely related to magnetic reconnection \citep{chen21, xu24, yang26a}. 
Two primary drivers have been identified in previous observational studies: flux emergence and flux cancellation, which correspond to 
two distinct models. In the emerging-flux model proposed by \cite{shibata92}, a magnetic bipole emerging from the low solar atmosphere 
or from below the photosphere interacts with the ambient pre-existing open magnetic field, triggering magnetic reconnection that 
ejects a jet along newly reconnected open field lines. This scenario has been supported by subsequent two-dimensional (2D) 
\citep{yokoyama95, yokoyama96} and 3D \citep{moreno08, moreno13} magnetohydrodynamics (MHD) simulations. Alternatively, \cite{sterling15} 
proposed another model. In their scenario, magnetic reconnection between open and closed fields was not driven by the emergence of a 
magnetic bipole, but by the eruption of a small-scale filament, with the photospheric magnetic signature in the source region being 
flux cancellation rather than flux emergence. The model was later simulated in 3D by \cite{wyper17}. Both models have received 
considerable observational support \citep{chen08, shen17, zheng18, duan24}, though the classical emerging-flux model is more applicable 
to standard jets, whereas the mini-filament-eruption model better accounts for the so-called blowout jets \citep{moore10}. 

In many observed cases, flux emergence is subsequently followed by flux cancellation, with associated jets often occurring during the 
cancellation phase \citep{tian17, tian18}. It is therefore expected that the occurrence and intensity of solar jets should be closely 
modulated by the evolution of the magnetic field in their source regions. Early on, \cite{roy73} proposed that significant magnetic 
flux changes in the source region are important for producing surges. However, quantitative studies on this relationship remain limited. 
On larger spatial and temporal scales, \cite{li12} statistically studied 575 solar jets associated with the 12-day evolution of AR 11045. 
They found that 15.3\%, 13.2\%, and 71.5\% of the jets occurred during the flux emergence, stable, and decaying phases, respectively. 
Recently, \cite{mulay16} analyzed 20 AR jets and reported that 50\% originated in flux cancellation regions, while 30\% occurred in 
flux emergence regions. Nevertheless, because the jets in their sample originated from different source regions, these statistical 
approaches are inherently unable to investigate the relationship between jet occurrence and magnetic field evolution in a single source 
region, let alone the relationship between jet intensity and magnetic flux change.

Beyond their axial propagation, solar jets frequently exhibit complex transverse motions, including drift of the jet axis 
\citep{chandrashekhar14}, lateral expansion of the jet spire \citep{shen11}, untwisting motion \citep{yang23b}, and transverse 
oscillations \citep{cirtain07}. These transverse motions provide crucial insights into the magnetic field configuration and reconnection 
process during jet ejection. Among them, drift motion, i.e., the overall sideways displacement of the jet axis, has received particular 
attention in both observations and modeling since early studies. Observations revealed that many jets exhibit overall sideways 
displacements of their axes \citep{canfield96}, with transverse velocities averaging several tens of km s$^{-1}$ \citep{shibata92, 
savcheva07}. Numerical simulations \citep{yokoyama95, yokoyama96} suggested that magnetic energy released via reconnection drives the 
jet, while the relaxation of the magnetic field after reconnection induces a sling-shot process that straightens the reconnected field 
lines under magnetic tension, naturally causing the plasma adhering to them to exhibit the observed drift motions, i.e., whip-like 
motions. Therefore, whip-like motion was regarded as strong evidence of magnetic reconnection.

On 2020 November 11, a group of homologous jets occurred in NOAA AR 12781 and were captured by the New Vacuum Solar Telescope 
\citep[NVST;][]{liu14} at Yunnan Observatory. Similar to homologous flares, homologous jets (or recurrent jets) are repetitive jet 
eruptions originating from the same source region, and thus can provide more detailed information about the magnetic topology of that 
region. The simultaneous line-of-sight (LOS) magnetograms provided by the Helioseismic and Magnetic Imager \citep[HMI;][]{scherrer12} 
on board the {\it Solar Dynamics Observatory} \citep[{\it SDO};][]{pesnell12} revealed that a magnetic bipole emerged in the source 
region just before the launch of the homologous jets, and its negative polarity subsequently began to cancel with the surrounding 
positive magnetic field. By continuously tracking the magnetic flux evolution of the bipole throughout its lifetime, we identify a 
clear three-phase evolutionary scenario with distinct jet responses: (1) during the flux-emergence phase, jets occur frequently but 
are relatively weak; (2) when the emerging flux enters its quasi-stable phase, jet activity drops to a minimum in both occurrence 
rate and intensity; and (3) during the flux-decay phase, both the occurrence rate and intensity of jet activity recover markedly, 
with the final jet being the most intense among all the jets and exhibiting a remarkable transverse drift motion. Although
photospheric magnetic flux decay can result from several different mechanisms, flux cancellation is the dominant process responsible 
for the observed decay in this event (see the Conclusions and Discussion section for the supporting arguments). We therefore refer 
to stage (3) as the flux-cancellation phase in the remainder of this paper. The drifting eruption of the final jet, occurring at the 
very end of the emerging bipole's life, provides clear evidence linking late-stage flux cancellation to the strongest energy release 
in the homologous jets. The remainder of this paper is organized as follows. Section 2 describes the instruments and data we used. 
Section 3 presents the detailed evolution of the jets and the photospheric magnetic field in the source region. Our conclusions and 
a brief discussion are presented in Section 4.

\section{Instruments and Methods}\label{sec:instru}
This event was well captured by NVST. NVST is a solar telescope located at the Fuxian Solar Observatory of Yunnan Observatories, 
with an aperture of 985 mm. It observes fine structures in the solar photosphere and chromosphere and provides high-resolution 
data in the $H\alpha$, He I and TiO wavebands. On 2020 November 11, NVST focused on NOAA AR 12781 through the $H\alpha$ line center 
(6562.8 \AA), its two off-bands ($\pm 0.6$ \AA), and the TiO channel. Here, we use only the $H\alpha$ line center and blue wing 
(-0.6 \AA) data to analyze the homologous jets. The NVST $H\alpha$ data have a field of view (FOV) of $180^{''} \times 180^{''}$, a 
pixel size of about $0^{''}.165$, and a time cadence of 44 s. The data were processed by dark-current subtraction and flat-field 
correction, and then reconstructed using the speckle masking method \citep{xiang16, cai22}.

We also use the level 1.5 full-disk images from the Atmospheric Imaging Assembly \citep[AIA;][]{lemen12} on board {\it SDO} to study 
the jets, and the full-disk LOS magnetograms obtained by HMI on board the same satellite to investigate the photospheric magnetic 
field in the source region of the jets. AIA takes images at 7 EUV passbands and 2 ultraviolet (UV) passbands, with a pixel size of 
$0^{''}.6$. The EUV images are used to identify the morphological features of the jets in this study, while the images at one UV 
channel (1600 \AA) are employed to investigate the jet bright points (JBPs). The time cadence is 12 s for AIA EUV images and 24 s 
for its UV images. The HMI LOS magnetograms have a pixel size of $0^{''}.5$ and a time cadence of 45 s.
 
All SDO data are processed using standard software programs in the SolarSoftWare ({\it SSW}) package and differentially rotated to 
a reference time close to the event. In addition, the NVST images are mapped to heliographic coordinates by registering them with 
SDO observations using an automatic mapping technique \citep{ji19, cai24}.

\section{Observational Results}\label{sec:result}
On 2020 November 11, NVST captured a series of solar jets in AR 12781. \cite{duan23} analyzed five groups of them and suggested 
that they are macrospicules whose formation mechanism is the same as that of coronal jets. In this paper, we investigate in detail 
a group of homologous jets included in the sample of \cite{duan23} but focus on different topics. Figure 1 shows four
representative jets among them in NVST $H\alpha -0.6$ \AA\ images. These jets all appear as dark features in $H\alpha$, with
different intensities and appearances. In particular, the last jet in this homologous group, J4, exhibits an obvious transverse 
drift motion, as shown in the bottom row of Figure 1. A careful inspection of the NVST $H\alpha -0.6$ \AA\ image sequence 
(see the animation accompanying the figure) shows that J4 was first observed to erupt at about 08:29:32 UT. By 08:36:11 UT (panel 
(d1)), it had developed a relatively distinct spire oriented almost along the white solid line in that panel. At about 08:39:52 UT 
(panel (d2)), the jet spire approximately reached its maximum, and the jet axis moved slightly southeastward, along the white 
dotted line shown in panel (d2). The dotted line is almost parallel to the solid line, indicating that the movement direction of 
the jet spire is almost perpendicular to its axis. The jet plasma began to fall back afterward, while the drift motion continued 
(panel (d3)). Notably, unlike the case of \cite{shen11}, the width of the jet spire did not change obviously during the lateral 
motion of its axis in this case. Therefore, the transverse motion of J4 is a lateral drift of its axis without expansion. We note 
that some dark plasma continued to connect the jet spire and the source region during its transverse movement, reminiscent of the 
whip-like motion of dark surges predicted by previous models and simulations \citep{shibata92, yokoyama95, yokoyama96}. Besides the 
drift motion, the animation accompanying Figure 1 reveals that J4 is the most intense among these homologous jets.

Figure 2 shows the general appearance of AR 12781 and the jet J4. The images in the top row of Figure 2 display an overview of the 
AR, while the images in the bottom row, which correspond to the region indicated by the blue rectangles in the top row, show a 
close-up view of J4 near its maximum and its magnetic environment. On 2020 November 11, AR 12781 was located at S$23\degr$W$35\degr$ 
with a typical $\beta$ configuration. The HMI magnetogram (panel (b1)) shows that the AR consists of a negative preceding polarity 
and a positive following polarity, conforming to the typical hemispheric rule for active regions in the southern hemisphere during 
solar cycle 25 \citep{hale19, xu22}. The HMI intensity image (panel (a1)) further reveals that the preceding polarity corresponds 
to a sunspot, while in the region of the following polarity, only some dispersed small pores are observed. In the corona, several 
dense loops are observed to connect the two main polarities of the AR (panel (d1)). The close-up images in the bottom row show that 
the source region of the homologous jets is located in the positive polarity region of the AR (panel (b2)), and that J4 appeared 
mainly as a dark collimated structure in both $H\alpha$ and EUV wavelengths when the spire reached its maximum. Notably, even with 
the high spatial resolution of NVST, no mini-filament was identified in the source region before the launch of J4 or any other jet 
in this homologous group. Instead, as we show below, a magnetic bipole was observed to emerge in the source region and cancel 
with the surrounding field. Therefore, the trigger of the homologous jets may be flux emergence rather than mini-filament eruptions. 

Among these homologous jets, J4 is the most distinctive owing to its strongest eruption and obvious drift motion. Therefore, we
investigate J4 in detail, and the results are presented in Figure 3 and the accompanying animation. The top and middle rows of 
Figure 3 show time-slice plots based on NVST $H\alpha$ line-center and AIA 304 \AA\ images, respectively. The long white arrows 
in the left panel of each row indicate the positions of two slits, S1 and S2. Notably, the slits are almost perpendicular to the 
spire of J4. In the time-slice plots, the drift motion of J4 appears as inclined dark bands, with the dotted lines approximately 
depicting their tendency. The plots show that the direction of J4's drift motion is opposite to the directions of arrows S1 and S2; 
thus, the jet spire drifted southeastward, or rotated clockwise when viewed from the JBP. In particular, we find that the widths of 
the dark bands do not show any significant increase over time, indicating that the jet spire drifted as a whole rather than expanding 
laterally \citep{shen11}. The velocities of the drift motion can be estimated approximately by calculating the slopes of these dark 
bands, and the resulting values are indicated above the bands. Overall, the drift velocities do not differ much when measured at 
different positions and wavelengths, but the values appear slightly larger when measured from $H\alpha$ images than from EUV images. 
The velocities range from $5.67$ km s$^{-1}$ to $7.88$ km s$^{-1}$ for the positions and wavelengths chosen here, which are slightly 
smaller than but still comparable to the results for whip-like motion of surges driven by emerging flux \citep{shibata92, savcheva07}. 
In particular, \cite {chandrashekhar14} recently investigated the drift motion of two polar coronal hole jets, one of which showed 
a drift motion with a velocity of $7.2$ km s$^{-1}$, in good agreement with our result. They suggested that this drift motion is 
caused by the progressive shift of the reconnection site along a sigmoid. Another characteristic of J4 is that the hot component 
is hardly identified throughout the lifetime of the jet in the NVST $H\alpha$ images, whereas in AIA EUV observations, hot and cool 
plasma coexisted in the jet spire early in the evolution of J4; the hot component then faded away rapidly while the cool component 
continued to enhance. The bottom row of Figure 3 presents the coexistence of the hot and cool components in some higher-temperature 
AIA EUV wavelengths, with the short black arrows indicating the cool component and the white arrows pointing to the hot component. 
This result is consistent with many previous studies, which revealed that hot and cool components are often observed simultaneously 
in a jet \citep{canfield96, joshi20}, or occur one after another in a short period \citep{jiang07, chen08, shen17}. According to 
numerical simulations \citep{yokoyama95, moreno08, li23}, the hot component is plasma heated by magnetic reconnection between the 
emerged closed field and the pre-existing open field (plasma at the reconnection site can be heated to X-ray temperatures), while 
the cool component originates directly from the emerging flux and is ejected along field lines different from those followed by the 
hot component due to the sling-shot effect.

The evolution of the photospheric magnetic field in the source region of the homologous jets is presented in Figure 4. As mentioned 
above, the source region was located in the positive polarity region of AR 12781, and the red rectangle in panel (b2) of Figure 2 
indicates the FOV of the HMI magnetograms shown in Figure 4 (panels (a1)-(a8)). From these magnetograms, we find that a magnetic 
bipole first emerged in the source region at 06:42:00 UT (panel (a2)). The positive polarity of the bipole, `p', is marked by the 
short white arrows in the figure, while the negative polarity, `n', is indicated by the short black arrows. Consistent with the 
previously reported motion of emerging bipoles \citep{jiang07}, p and n showed a separating motion, and the intensities of both 
polarities increased. However, in this event, the movements of p and n were not a simple separation along their connecting line. 
Perhaps because it was hindered by the surrounding strong magnetic field of the same polarity, p hardly moved during the emergence 
process, whereas n moved continuously toward the northwest. As a result, the axis of the emerging bipole rotated clockwise. By drawing 
in panel (a6) a red dashed line for the initial axis position (from panel (a2)) and a blue dashed line for its position 
at 08:55:30 UT (from panel (a8)), we find that the axis of the emerging bipole rotated approximately $40.6\degr$ over 133 minutes.

Following the emergence and movement of the negative polarity, n, magnetic reconnection is expected to take place between this 
polarity and the pre-existing ambient positive field, which would trigger coronal jets according to previous models and simulations 
\citep{shibata92, yokoyama95}. It is known that solar jets are often associated with the movements of satellite spots or moving 
magnetic features \citep{canfield96, chen15}. This also applies to our event, in which the homologous jets are associated with the 
movement of a magnetic bipole. The HMI observations show that the negative polarity of the bipole, n, moved northwestward, 
approximately along a trajectory indicated by the long blue arrow S3 in panel (a4); we therefore constructed a time-slice plot 
along this slit to depict its movement (panel (b)). The time-slice plot indicates that the movement of n is almost linear 
(indicated by the white dashed line in panel (b)), with a uniform velocity of about $0.64$ km s$^{-1}$, estimated from a linear 
fit to the time-slice. In a study of recurrent and homologous AR jets, \cite{chen15} found that an associated satellite spot moved 
with an average velocity of $0.29$ km s$^{-1}$, and that a moving magnetic feature moved at a velocity of $0.48$ km s$^{-1}$. Both 
values are consistent with the velocity of n in our case. 

Almost simultaneously with the emergence of the bipole, a pair of brightenings appeared in AIA 1600 \AA\ images and showed 
separating motion also (Figure 5). Similar phenomena have been reported by \cite{shen22}. These brightenings likely provide 
evidence of magnetic reconnection between the emerging bipole and the pre-existing coronal magnetic field. According to the 
emerging-flux model of \cite{shibata92}, the northern brightening (indicated by the short black arrow in panel (a2)) corresponds 
to newly formed closed loops after magnetic reconnection between n and the surrounding positive field, or, from an observational 
perspective, to the JBPs of the homologous jets, whereas the southern brightening (indicated by the white arrow in panel (a2)) is 
the footpoint of the homologous jets rooted in the emerged positive polarity p. Therefore, it is no surprise that the brightenings 
show movement trajectories similar to that of the bipole: the southern brightening remained almost at its initial position, while 
the northern brightening moved northwestward. The continuous motion of the northern brightening (the JBPs) may suggest that the 
magnetic reconnection site was also sliding, even though the brightening and the reconnection site are not co-spatial in the solar 
atmosphere. Panel (b) presents the corresponding time-slice plot along slit S3. As indicated by the long blue arrows in panels (a5) 
and (a7), S3 has the same projected position as the slit used in Figure 4 but samples a different atmospheric layer. The trajectory 
of the JBPs appears as an inclined bright band in the time-slice plot, with an estimated average velocity of $0.80$ km s$^{-1}$.

Based on Figures 4 and 5, we find that the negative polarity of the bipole moved toward the northwest and underwent continuous 
reconnection with the surrounding positive field after it emerged into the photosphere. Consistently, NVST $H\alpha$ and AIA EUV 
observations revealed that a group of homologous jets erupted from this region throughout the period. However, as shown in Figure 1, 
the intensities of these homologous jets vary greatly. To investigate the relationship between jet intensity and the evolution of 
magnetic flux in the source region, we present in Figure 6 a time-slice plot based on NVST $H\alpha$ blue wing images, overlaid with 
the HMI LOS negative flux (absolute value) and the AIA 1600 \AA\ intensity as blue and white curves, respectively. The time-slice 
was constructed along a slit SS, shown by a long white arrow in panel (a) of Figure 1, while the magnetic flux and the 1600 \AA\ 
intensity were calculated in a region indicated by the red ovals in Figures 4 and 5. The time-slice plot shows that many jets ejected 
along the slit SS between 06:30 UT and 09:00 UT, and the four representative jets shown in Figure 1 (J1-J4) are marked by yellow 
vertical arrows. The blue curve shows that the evolution of the emerging bipole's negative flux can be clearly divided into three 
stages: the flux-emergence stage from 06:42 UT to 07:18 UT, the approximately flux-stable stage from 07:18 UT to 07:54 UT, and the 
flux-cancellation stage after 07:54 UT. Combining the time-slice plot and the flux curve, it is noticed that during the flux-emergence 
stage, jet activity is frequent and of moderate intensity; during the flux-stable stage, jet activity is relatively weak in both 
frequency and intensity (e.g., J2); and strong jet activities (J3 and J4) occur during the flux-cancellation phase. In particular, 
the strongest one among these homologous jets, J4, occurred in the late stage of flux cancellation. Because the emerging negative
polarity n moved continuously, both the magnetic reconnection site and the jet footpoint moved accordingly. Consequently, the initial 
trajectory of J4 deviated from the slit SS, resulting in the absence of J4's initial eruption phase and an apparent gap between J4 and 
the other jets in Figure 6. Checking the NVST $H\alpha$ blue wing observations carefully (see the animation accompanying Figure 1),
we infer that J4 erupted no later than 08:29:32 UT, a time indicated by a vertical red dotted line in Figure 6. On the other hand, 
the AIA 1600 \AA\ light curve (white curve) shows the intensity variations of the JBPs. The fluctuations of the curve represent the 
appearance and disappearance of JBPs corresponding to different jets. It is obvious that the highest peak of the light curve was 
associated with J4, indicating that the strongest jet was accompanied by the brightest JBP in the sequence. The brightness of 
this brightest JBP began to increase almost simultaneously with the negative magnetic flux reaching its minimum, about two minutes 
before the eruption of J4. Therefore, J4 occurred right after the continuous and rapid cancellation of the magnetic flux. Similar 
results have been reported previously for jets triggered by mini-filament eruptions \citep{panesar16, yang24}, and \cite{sterling18} 
proposed that a substantial fraction of the region's total flux may need to be cancelled so that enough magnetic free energy 
can accumulate in the system to trigger a jet. Overall, we infer from the figure that jet activity during the flux-cancellation 
stage is stronger than that during the flux-emergence and flux-stable stages, and that the strongest activity occurred in the late 
stage of flux cancellation.

\section{Conclusions and Discussion}\label{sec:discussion}
In this study, we investigate the relationship between the evolutionary phases of an emerging magnetic bipole and the properties of 
the homologous jets it triggers. Our main findings are summarized as follows:

\begin{enumerate}
\item The temporal evolution of the negative flux of the bipole is characterized by emergence, stabilization, and cancellation stages. 
The resulting jets exhibit distinct behaviors at each stage: frequent but mild eruptions during the emergence phase, minimal activity 
during the stable phase, and a significant resurgence with intensified eruptions during the cancellation phase. The most violent jet 
in the entire group of homologous jets, J4, was ejected a few minutes after the rapid decrease of the negative flux, implying that 
the most intense release of magnetic free energy occurs at the late stage of flux cancellation.

\item Since its emergence, the negative polarity of the bipole moved continuously northwestward at a velocity of $0.64$ km s$^{-1}$ .
Accordingly, the JBPs of the homologous jets moved in the same direction at a velocity of $0.80$ km s$^{-1}$, indicating sliding
of the reconnection site.

\item The dominant jet, J4, is distinguished not only by its high eruption intensity but also by the brightest JBP emission and a 
clear lateral drift of its axis. The measured drift velocities ranged from $5.67$ km s$^{-1}$ to $7.88$ km s$^{-1}$ in NVST $H\alpha$ 
and AIA 304 \AA\ observations. The drift motion of J4 might result from the combined effect of the sling-shot process and sliding 
magnetic reconnection.

\end{enumerate}

Homologous solar coronal jets have occasionally been reported in previous observational studies, but few studies have focused on the 
relationship between the properties of homologous jets and the evolution of the magnetic flux in their source region. Our study clearly 
shows that the number and intensity of the resulting jets are closely correlated with the magnetic flux in the source region, indicating 
that the efficiency of magnetic reconnection between the emerging bipole and the pre-existing ambient field is modulated by the state 
of flux evolution. Previous studies of individual jet events triggered by emerging flux found that jets are more likely to be ejected 
during the flux-cancellation stage shortly after flux emergence \citep{li12, mulay16, tian17}. However, this study shows that many jets 
occurred during the flux-emergence stage, although these eruptions are not as violent as the jets triggered during the flux-cancellation 
stage. This can be interpreted as follows: magnetic reconnection took place as soon as the bipole emerged and triggered the jets, but 
magnetic free energy had not accumulated sufficiently during this stage, thus the jets at the flux-emergence stage are mild. In contrast, 
at the late stage of flux cancellation, a large amount of magnetic free energy had accumulated in the magnetic system during the preceding 
emergence, stabilization, and rapid cancellation phases, and was suddenly released to drive the strongest eruptions. On the other hand, 
the jets triggered during the flux-stable stage were the fewest and weakest among the three stages, indicating that magnetic reconnection 
between the emerged bipole and the ambient field was weak. Therefore, during the flux-stable stage, flux emergence and cancellation may 
nearly cease rather than reach equilibrium.

Although we loosely refer to the third stage of the negative flux evolution in this event as the flux-cancellation phase, strictly 
speaking it is a flux-decay phase, because a decrease in magnetic flux is not necessarily caused by flux cancellation. According to 
\cite{parker84}, three mechanisms can remove magnetic flux from the photosphere: bodily transport of magnetic flux out of the region, 
cancellation of opposite-polarity flux, and bodily submergence of an entire flux bundle without reconnection. It is widely accepted that
the first mechanism contributes significantly to flux removal in ARs. However, in the present event, the HMI LOS magnetograms in Figure 5 
show that the bulk of the emerging negative polarity n was not transported out of the red-oval region used for the magnetic flux  
calculation. \cite{kutsenko24} presented evidence for the third mechanism in a study of flux removal in decaying ARs. They found that,
during the flux-decay process of each AR they studied, the main opposite-polarity sunspots connected by coronal loops gradually approached 
each other, but no reconnection signatures such as flare-like bursts in UV emission were observed. They therefore suggested that the 
convergence was induced by the submergence of an AR-size loop. In our case, the two magnetic polarities of the emerging bipole have 
continued to move apart since their emergence, which is inconsistent with the converging motion expected for loop submergence. Instead, 
we note that the separating motion of p and n drives n to gradually approach the surrounding pre-existing positive polarities, and that 
the jet eruptions and the appearance of 1600 \AA\ brightenings (the JBPs) indicate magnetic reconnection between n and the pre-existing 
positive field. We therefore suggest that flux cancellation is the dominant process responsible for the flux decay in the source region 
of the jets studied here, even if other mechanisms cannot yet be completely ruled out.

The production of solar jets is thought to be closely correlated with moving satellite sunspots, magnetic bipoles, or moving magnetic 
features \citep{yang26b}. According to \cite{canfield96}, moving satellite sunspots or magnetic bipoles could create conditions favorable 
to magnetic reconnection and thus to triggering jets. \cite{chen15} analyzed recurrent and homologous jets occurring at the west edge 
of an AR and found that they are induced by collisions of a satellite spot and moving magnetic features with the pre-existing opposite 
polarity magnetic field. \cite{yang13} presented a 2.5 dimensional numerical simulation of chromospheric anemone jets triggered by 
reconnection between moving magnetic features and the pre-existing ambient field. In our case, the negative polarity of the emerging 
magnetic bipole continuously moved northwestward and cancelled with the pre-existing surrounding positive field, triggering the homologous 
jets. However, a special feature of the emerging bipole in our case is that its positive polarity remained almost stationary while the 
negative polarity moved. As a result, the axis of the emerged bipole rotated clockwise by $40.6\degr$ over 133 minutes (see Figure 4). 
According to previous studies, sunspot rotation (including a sunspot rotating around its umbral center and a sunspot rotating around 
another sunspot) is regarded as related to flare activity and filament formation/eruption \citep{yan08, yan16}. As a miniature version 
of sunspots, the magnetic bipole in our case may promote the launch of the jets through its rotation, although current observations 
are not sufficient to confirm this. Furthermore, the rotation of the bipole in our case is much faster than that of sunspots. For example, 
\cite{yan16} reported that a small positive sunspot rotated around a negative sunspot by about $40\degr$ over 33 hr. The higher 
rotational velocity of the bipole is most likely due to its smaller volume compared with sunspots.
 
The major jet J4 in these homologous jets showed significant lateral motion almost perpendicular to the jet spire. Both NVST $H\alpha$ 
and AIA EUV observations showed that this lateral motion was simply the drift of the jet axis, without either expanding or untwisting
of the spire. In the emerging-flux model, the drift motion of the jet axis was interpreted as the whip-like motion induced by the 
sling-shot process of magnetic reconnection \citep{shibata92, canfield96}. \cite{savcheva07} presented the first statistical 
observational study of the drift motion, and the transverse velocities they obtained were in the range of 0–35 km s$^{-1}$. The 
simulations of \cite{yokoyama95, yokoyama96} reproduced the whip-like motion of the cool plasma, with an estimated velocity of 48-84 
km s$^{-1}$. Another cause of the drift motion is the gradual change of the reconnection site. For instance, the 3D simulations of 
\cite{moreno08, moreno13} revealed that a split-vault with two chambers formed below a jet, containing the emerged and reconnected 
field loops respectively. As magnetic reconnection proceeded, the chamber containing the emerged loops shrank while that containing 
the reconnected loops grew, thus resulting in an apparent horizontal drift of the vault and the jet. The estimated velocity of this 
drift motion is roughly 10 km s$^{-1}$. Notably, the split-vault with two chambers was confirmed by the observations of \cite{joshi20} 
at the west limb. The observational study of \cite{chandrashekhar14} revealed that the drift motion of two jets was due to the 
progressive shift of the magnetic reconnection site along a magnetic structure (a micro-sigmoid) at the jet base, and that the length 
of the transverse motion is similar to that of the JBP. Therefore, they proposed that the jets were produced by a series of small-scale 
magnetic reconnections occurring at different times and locations. In our case, HMI LOS magnetograms show continuous motion of the 
negative polarity of the emerging bipole, while AIA 1600 \AA\ images show corresponding motion of the JBP. Both indicate that the 
reconnection site was moving, suggesting that its sliding may also contribute to the drift motion of J4. However, we note that the 
negative polarity of the bipole and the JBP moved northwestward at a nearly constant speed throughout the homologous jets sequence, 
whereas the other jets in the group did not exhibit drift motion as obvious as that of J4. Moreover, the velocities of the negative 
polarity and the JBP are only about 10\% of J4's drift velocity. Taken together, these two observations lead us to propose that the 
drift motion of J4 results from the combined effect of the whip-like motion induced by the sling-shot process and the sliding of the 
reconnection site, with the former being the dominant factor and the latter secondary. This is consistent with the fact that the 
magnetic free energy released during the earlier jets in the group is less intense than that during J4, so the sling-shot effect and 
the associated whip-like motion are weaker for the other jets. Although the motions of the negative polarity and the JBP indicate 
that the reconnection site was also moving, they do not necessarily trace the direction of its motion. Current observations cannot 
determine the sliding direction of the reconnection site, thus we cannot tell whether it promoted or hindered the whip-like motion of J4. 

Overall, this study suggests that homologous jets may serve as probes of the ongoing evolution of the magnetic field in the low solar 
atmosphere, and provide critical insights into the recurring release of magnetic free energy. The sequential accumulation and abrupt 
release of magnetic energy during flux emergence and cancellation offer clues for understanding the intermittency and intensity 
variations of recurring solar jets.

\begin{acknowledgments}
We thank an anonymous referee for the constructive suggestions that helped remedy omissions and improve the quality of this paper.
We thank the NVST and the {\it SDO} teams for granting free access to their internet databases, and thank Dr. Xu Zhe for the 
assistance provided in NVST data processing. This work is supported by the Strategic Priority Research Program of the Chinese Academy 
of Sciences, Grant No. XDB0560000, the National Natural Science Foundation of China (12273108, 12673065 and 12273106), the Yunnan Science 
Foundation of China (202601AT070305 and 202501AT070025), the Yunnan Province XingDian Talent Support Program, and by the CAS “Light of 
West China” Program.
\end{acknowledgments}

\bibliography{ms}{}

\clearpage

\begin{figure}
\plotone{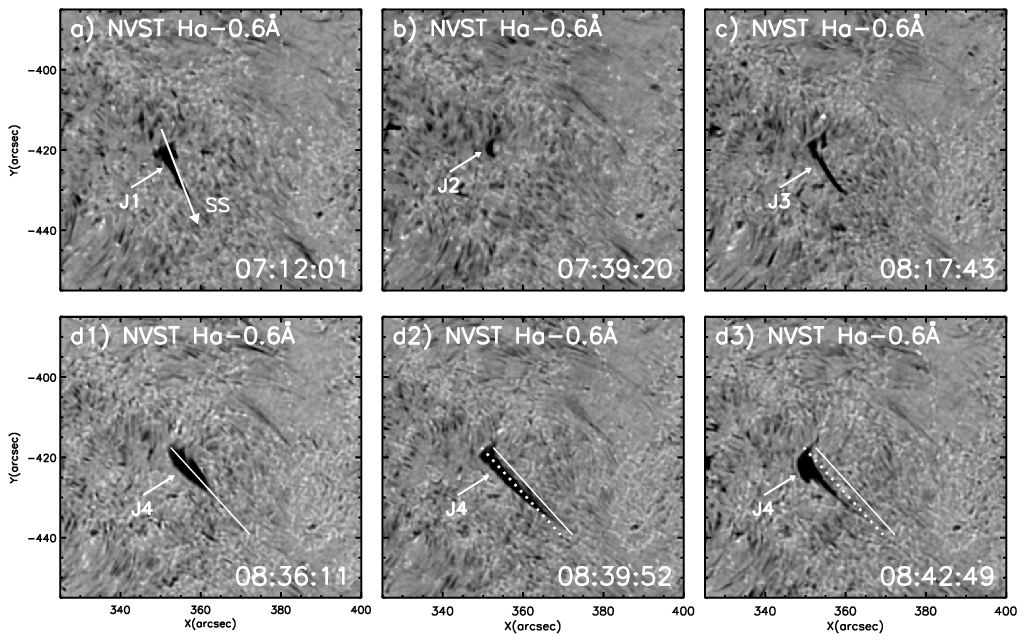}   
\caption{NVST H$\alpha -0.6$ \AA\ images showing four representative jets, J1-J4, among the homologous jets that occurred in 
AR 12781. The bottom row demonstrates the lateral drift motion of J4, where the solid lines indicate its position at 08:36:11 UT 
and the dotted lines indicate its position at 08:39:52 UT. The long arrow in panel (a) indicates the position of a slit, SS,
along which a time-slice plot was constructed and is presented in Figure 6. (An animation of this figure is available, with the 
same FOV as the figure and a time cadence of about 45 s. It lasts 20 s and covers the period from 06:30 UT to 09:00 UT on 2020 
November 11.)
\label{fig1}}
\end{figure}

\begin{figure}
\epsscale{0.95}
\plotone{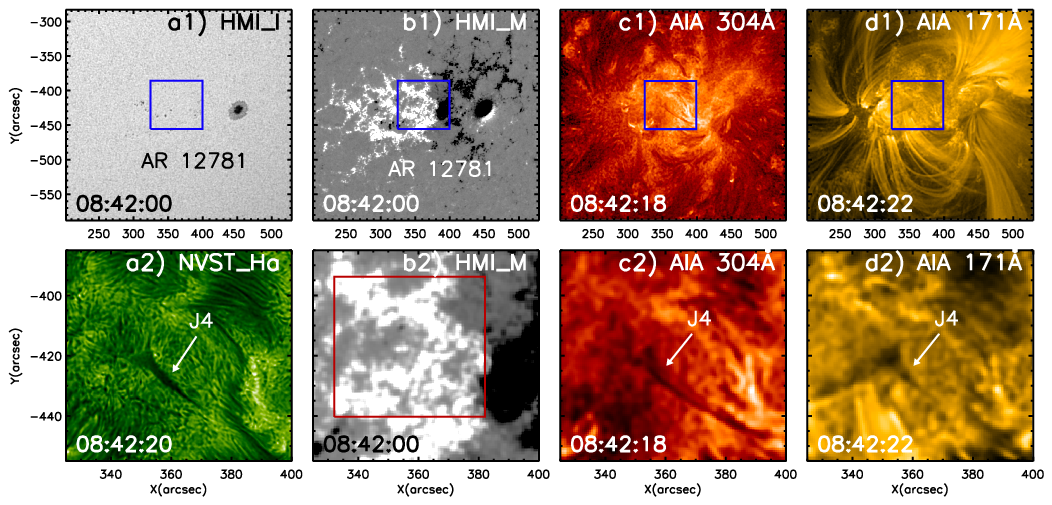}   
\caption{General appearance of AR 12781 (top row) and the jet J4 (bottom row). The blue rectangles in the top row indicate the 
FOV of the images in the bottom row, which is the same as that of Figures 1 and 3. The red rectangle in panel (b2) shows the FOV 
of the magnetograms and AIA 1600 \AA\ images in Figures 4 and 5. The white arrows indicate J4 when it is approximately at its maximum. 
\label{fig2}}
\end{figure}

\begin{figure}
\epsscale{0.95}
\plotone{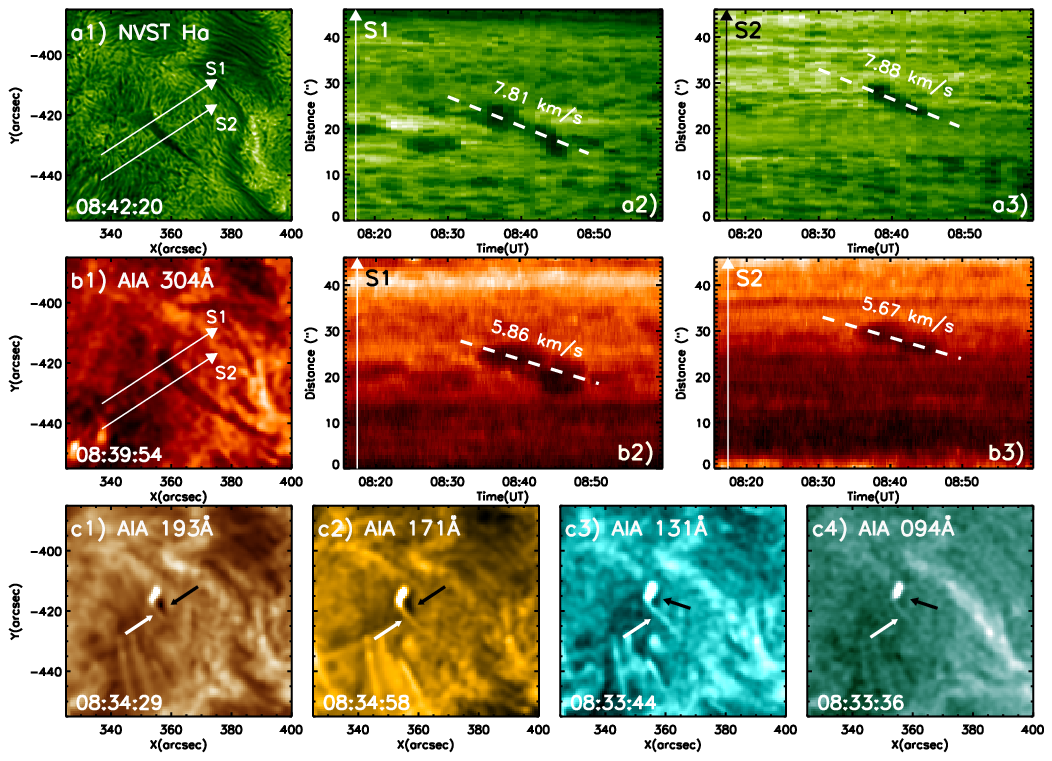}   
\caption{Lateral drift motion of jet J4 depicted by time-slice plots at NVST H$\alpha$ line center (top row) and SDO/AIA 304 \AA\ 
wavelength (middle row), and the hot and cool components of J4 shown by AIA multi-channel images (bottom row). The positions of 
slits S1 and S2 are shown in panels (a1) and (b1) as two long arrows. The slanted white dashed lines in the time-slice plots 
indicate the movement of J4's cool component, with the estimated velocities shown above them. The short white and black arrows in 
the bottom row indicate the hot and cool components of J4, respectively. (An animation of NVST H$\alpha$ line center images and 
AIA 304, 171, and 94 \AA\ direct images accompanies this Figure. The animation has the same FOV as the Figure, with a time cadence 
of about 45 s for NVST H$\alpha$ images and 12 s for AIA images. It lasts 7 s and covers the period from 08:20 UT to 09:00 UT on 
2020 November 11.)
\label{fig3}}
\end{figure}

\begin{figure}
\epsscale{0.95}
\plotone{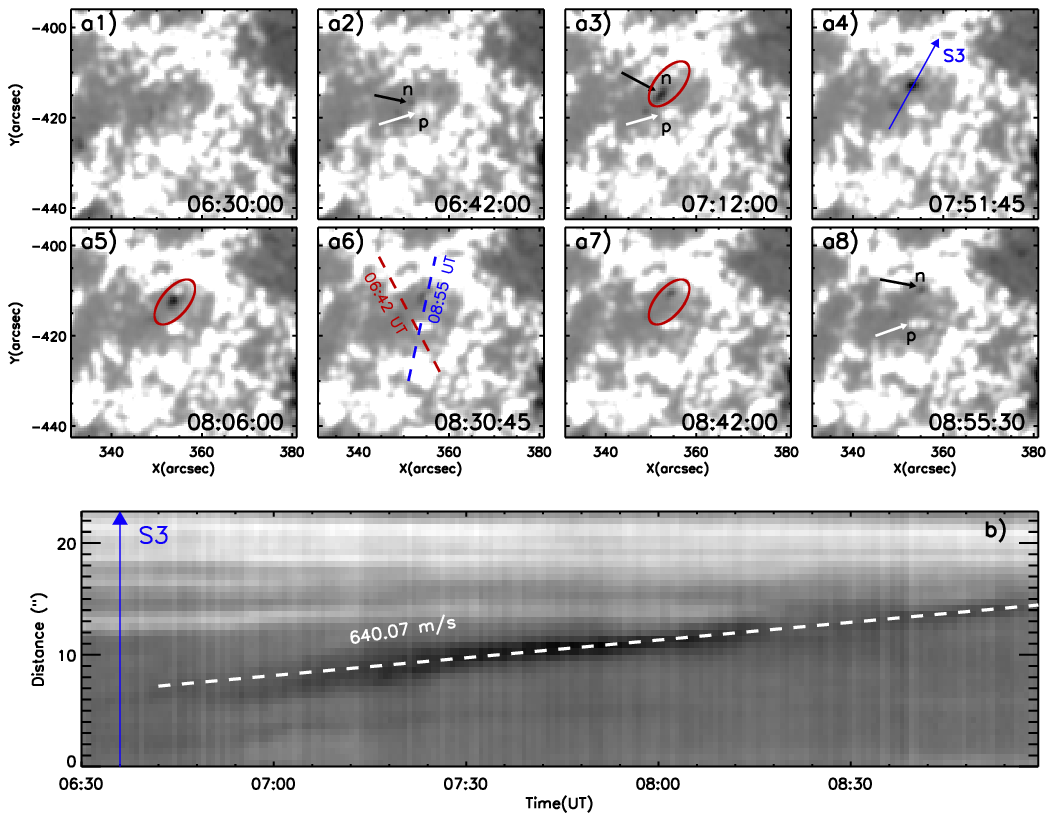}   
\caption{HMI magnetograms (a1-a8) and the corresponding time-slice plot (b) show the emergence and movement of the magnetic bipole 
in the source region of the homologous jets. The time-slice plot was constructed along slit S3, which is indicated by the long blue 
arrow in panel (a4). The short black and white arrows indicate the negative polarity `n' and the positive polarity `p' of the 
emerged bipole, respectively. In panel (a6), the red and blue dashed lines represent the p-n connecting lines at 06:42:00 UT and 
08:55:30 UT, respectively. The red ovals encompass the region over which n moved, where the magnetic flux was calculated and is 
presented in Figure 6. The FOV of the magnetograms is indicated by the red rectangle in panel (b2) of Figure 2.
\label{fig4}}
\end{figure}

\begin{figure}
\epsscale{0.95}
\plotone{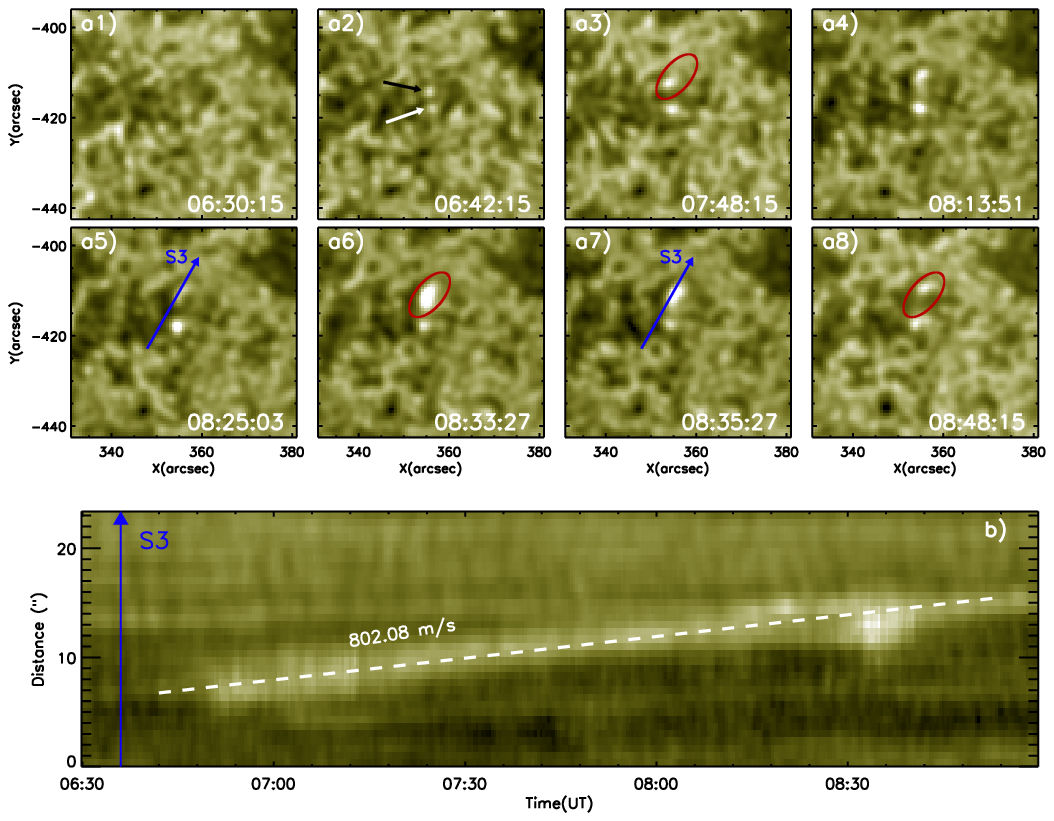}   
\caption{AIA 1600 \AA\ images (a1-a8) and the corresponding time-slice plot (b) show the evolution of the brightenings that 
appeared in the source region of the homologous jets along with the emergence of the magnetic bipole. The black and white arrows 
in panel (a2) indicate a pair of brightenings. The time-slice is constructed along slit S3 which is indicated by the long blue 
arrows, and the red ovals denote the region where the AIA 1600 \AA\ intensities were calculated and are presented in Figure 6. The 
white dotted line and the number above it in panel (b) indicate the movement of the northern brightening and its estimated velocity.
The positions of the long blue arrow and the red oval, as well as the FOV of the images, are the same as those in Figure 4.
\label{fig5}}
\end{figure}

\begin{figure}
\epsscale{0.95}
\plotone{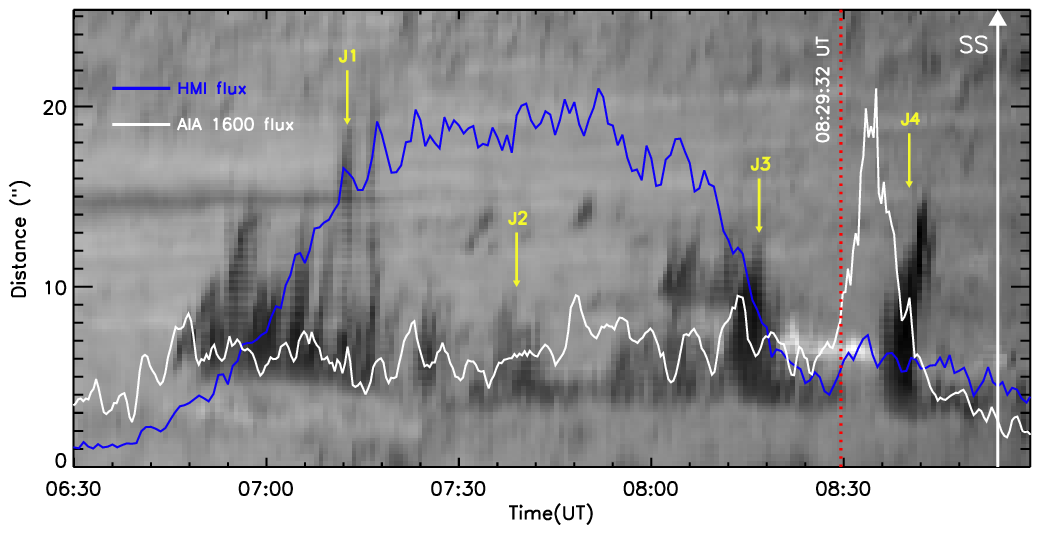}   
\caption{Relationship between the strength of the recurrent jets and the HMI magnetic flux. The background image is a time-slice 
plot based on NVST H$\alpha$ blue wing images, constructed along slit SS, which is indicated by the long white arrow in panel (a) 
of Figure 1. The overlying curves indicate variations in the HMI negative magnetic flux (blue) and the AIA 1600 \AA\ intensity (white). 
Both quantities are measured in the region marked by the red ovals in Figures 4 and 5. The long white arrow on the right side of the 
figure indicates slit SS, the four short yellow arrows indicate the four representative jets J1-J4 shown in Figure 1, while the 
vertical red dotted line markes the time when J4 first appeared in NVST H$\alpha$ blue wing images.
\label{fig6}}
\end{figure}

\end{document}